\documentstyle[12pt,frascatiphys]{article}
\begin{document}
\title{ 
LATEST RESULTS FROM LATTICE QCD FOR EXOTIC
HYBRID MESONS.
}
\author{
Craig McNeile        \\
{\em Dept. of Math Sci., University of Liverpool, L69 3BX, UK.} 
}
\maketitle
\baselineskip=14.5pt
\begin{abstract}
I review the results from lattice gauge theory for the 
masses of exotic hybrid mesons.
\end{abstract}
\baselineskip=17pt
\section{Introduction}
The quark model predicts that the 
charge conjugation (C) and parity (P) of a meson
with spin $S$ and orbital angular momentum $L$  is
\begin{equation} 
P  =  (-1)^{L+1} \; \; \;
C  =  (-1)^{L+S} 
\label{eq:JCP}
\end{equation} 
States with quantum numbers not  produced 
by eq.~\ref{eq:JCP}, 
such as
\begin{equation}
J_{exotic}^{PC} = 1^{-+}, 0^{+-}, 2^{+-}, 0^{--}
\end{equation}
are known as exotics~\cite{Burnett:1990aw}.  
Exotic states are allowed by QCD.

There are a number of different possibilities for the structure of an
exotic state. An exotic signal could be: a hybrid meson, which is
quark and anti-quark and excited glue, or bound state of two quarks and
two anti-quarks ($\overline{Q}\overline{Q}QQ$). The two most popular
guesses for the structure of the ($\overline{Q}\overline{Q}QQ$) state
are either a molecule of two mesons or diquark anti-diquark bound
state.  In this paper I review the latest lattice results for the
masses of exotic hybrid mesons, concentrating on the $1^{-+}$ state,
obtained from lattice QCD.

\section{Lattice simulations of exotic mesons}



Many numerical predictions of QCD can be 
determined from the path integral
\begin{equation}
c(t) \sim \int dU  \int d\psi  \int d\overline{\psi}
\;
\sum_{\underline{x}} O(\underline{0},0) 
O(\underline{x},t)^{\dagger}
e^{-S_F - S_G }
\label{eq:reallyQCD}
\end{equation}
where $S_F$ is the fermion action (some lattice
discretization of the Dirac action) and $S_G$
is the pure gauge action. 
The path integral in eq.~\ref{eq:reallyQCD} is put on
the computer using a clever finite 
difference formalism~\cite{Montvay:1994cy}, due 
to Wilson, that maintains gauge invariance.
The physical picture for eq.~\ref{eq:reallyQCD} is that a
hadron is created at time 0, from where it propagates to the time t,
where it is destroyed.
The fermion integration can be done exactly in
eq.~\ref{eq:reallyQCD} to produce the fermion determinant.
Simulations that include the effect of the determinant are very
expensive computationally, so typically it is 
not included in the simulation (the quenched approximation).
However there has been some recent work that includes the
effect of the determinant~\cite{Lacock:1998A} on the light
exotic spectrum.

The standard interpolating operator for the pion, which can be used 
in eq.~\ref{eq:reallyQCD},  is
\begin{equation}
O_{\pi} (\underline{x} , t ) = \overline{\psi}(\underline{x},t)
\gamma_5 
\psi (\underline{x},t)
\end{equation}
which has the correct $J^{PC} = 0^{-+}$ quantum numbers.
One possible interpolating operator~\cite{Bernard:1997ib} 
for an exotic $1^{-+}$ particle is
\begin{equation}
O_{1^{-+}} (\underline{x} , t ) = \overline{\psi}(\underline{x},t)
\gamma_j F_{ij} (\underline{x},t)
\psi (\underline{x},t)
\label{eq:interponemp}
\end{equation}
where $F$ is the QCD field strength tensor.
It is essential to use operators that have 
some spatial separation between the quarks in the meson
to get a good signal. Recently the MILC collaboration has
attempted to measure the ``wave function'' of the $1^{-+}$ hybrid meson
in coulomb gauge~\cite{McNeile:1998cp}. Unfortunately the operator
used did not have the correct charge conjugation quantum number, so 
the published wave function~\cite{McNeile:1998cp} is incorrect.

In this formalism a gauge invariant interpolating operator, for any
possible exotic hybrid particle or four particle state can be
constructed. The dynamics then determines whether the resulting state
has a narrow decay width, which can be detected experimentally. In the
large $N_c$ (number of colours)
limit~\cite{Burnett:1990aw,Cohen:1998jb} both exotic hybrid mesons and
non-exotic mesons have widths vanishingly small compared to their
masses.

The data from the simulation is extracted using a 
fit model~\cite{Montvay:1994cy}:
\begin{equation}
c(t) = a_0 exp( -m_0 t ) + a_1 exp( -m_1 t ) + \cdots
\label{eq:fitmodel}
\end{equation}
where $m_0$ ($m_1$) is the ground (first excited) state
mass and the dots represent higher excitations.
Although in principle excited state masses can be extracted from a
multiple exponential fit, in practice this is numerically non-trivial
because of the noise in the data from the simulation.
Simulations that involve the calculation of the properties
of exotic hybrid mesons are harder than those that concentrate
on the non-exotic hadrons, because the signal to noise ratio is 
worse for exotic mesons than for $\overline{Q}Q$ mesons.

In an individual lattice simulation there are errors from the finite
size of the lattice spacing and the finite lattice volume. State of the art
lattice simulations in the quenched theory, run at a number of
different lattice spacings and physical volumes and extrapolate the
results to the continuum and infinite volume. For the exotic mesons,
this is done for heavy quarks (see section~\ref{se:heavyRES}), but as
yet, the continuum extrapolation has not been done for light exotic mesons.


How successful is lattice QCD in practice? One way to tell is to
compare the predictions for the masses from lattice QCD for well known
particles (proton, $\rho$, etc.) with experiment.  The most accurate
quenched calculation to date has recently been completed by the
CP-PACS collaboration~\cite{Kanaya:1998sd}.  From the masses of 11
light hadrons, they conclude that the quenched approximation disagrees
with experiment by at most 11\%.  The comparison of results from
simulations that include dynamical fermions with experiment is less
clear, because of their high computational cost (see
Kenway~\cite{Kenway:1998ew} for a review of the latest results).

The results from lattice QCD also provide insight into the underlying
dynamics of light hadrons. Lattice QCD simulations can test the
various assumptions made in models of the QCD dynamics.  For example
there are a number of models of exotic states based on the idea of a
bound diquark anti-diquark
pair~\cite{Uehara:1996hc}.  A
critical assumption in diquark models is that two quarks actually do
cluster to form a diquark. This assumption has recently been tested in a
lattice gauge theory simulation by the Bielefeld
group~\cite{Hess:1998sd}, where they found no deeply bound diquark
state in Landau gauge.

\section{Results for light exotic mesons}

In the last year the MILC collaboration have
repeated~\cite{McNeile:1998cp} their initial
simulations~\cite{Bernard:1997ib} using the clover fermion action.
The clover action is ``closer'' to the continuum than
the Wilson fermion action, because it has the leading order lattice
spacing terms removed.  There are also new results for the hybrid
masses from the SESAM collaboration (reported by Lacock and Schilling)
~\cite{Lacock:1998A}, that include some effects from dynamical sea
fermions.

The results for the mass of the $1^{-+}$ exotic state are summarised in
table~\ref{onempRES}.  All the results from the various simulations
are essentially consistent.
with the mass of the $1^{-+}$ state  around $2$ GeV.
In table~\ref{describeLIGHT} we show the
physical parameters for each of the simulations. The interpolating
operators used to create the exotic meson states in the MILC
calculations~\cite{Bernard:1997ib} are different to those used in the
UKQCD~\cite{Lacock:1997ny} and SESAM simulations~\cite{Lacock:1998A}.

The observation that the results for the mass of the $1^{-+}$
hybrid meson are consistent for very 
different simulations gives us confidence in the final result.
Although I would prefer to see simulation results at lighter
quark masses. Simulations at the point where the ratio of the 
pseudoscalar mass to vector mass ($M_{PS}/M_V \sim 0.5  $) are
possible with current algorithms and computers~\cite{Kanaya:1998sd}.

\begin{table}
\centering
\caption{ \it Masses of the light $1^{-+}$ hybrid from
lattice gauge theory.
}
\vskip 0.1 in
\begin{tabular}{|c|c|c|} \hline
Simulation          &  Group & mass GeV \\
\hline
\hline
A & UKQCD~\cite{Lacock:1997ny}   & 
             $ 2.0(2)$                    \\
B & MILC~\cite{Bernard:1997ib,McNeile:1998cp}    &  $ 2.0(1) \pm sys $   \\
C & MILC~\cite{Bernard:1997ib,McNeile:1998cp}    &  $ 2.1(1) \pm sys $   \\
D & Lacock and Schilling~\cite{Lacock:1998A}    &   $ 1.9(2) $  \\
\hline
\end{tabular}
\label{onempRES}
\end{table}
%


\begin{table}
\centering
\caption{\it Parameters of light exotic meson simulations,
}
\vskip 0.1 in
\begin{tabular}{|c|c|c|c|c|c|} \hline
Simulation &   Action &  fermions & length
(fm) & $a^{-1}$ GeV & $M_{PS} / M_V$  \\
\hline
\hline
A   & clover   & quenched & $1.6$  & $2.0$  & 
  $0.76$  \\ 
B   & Wilson  &  quenched & $2.3$  & 
$2.8$ &  $0.96,0.93,0.88,0.77$  \\ 
C   & clover  &  quenched & $2.3$ & 
$2.8$ & $0.94,0.90,0.72$   \\ 
D   & Wilson  &  dynamical & $1.4$ & 
$2.3$ & $0.83,0.81,0.76,0.69$   \\ 
\hline
\end{tabular}
\label{describeLIGHT}
\end{table}
%


\section{Results for heavy exotic mesons} \label{se:heavyRES}

There has been a lot of work on calculating the spectroscopy of $\overline{c}c$
and $\overline{b}b$ mesons from lattice gauge theory 
(see Davies~\cite{Davies:1997hv} for a review). 
The main technical complication in heavy quark simulations is that the lattice
spacing of current simulations is not smaller than the heavy quark
mass. So various effective field theory Lagrangian approximations
to QCD are simulated.

The NRQCD (nonrelativistic QCD) Lagrangian is one such effective
field theory approximation to QCD, with the
expansion parameter equal to the velocity squared. 
NRQCD has been particularly successful
in simulating the Upsilon spectrum~\cite{Davies:1997hv},
but is less well converged for charmomium, (particularly for
spin splittings), because the charm quarks move with higher
velocity~\cite{Davies:1997hv}.
The NRQCD Langrangian
correct up to $O(Mv^2)$ is
\begin{equation}
{\cal L}^{NRQCD} = \overline{\psi}( -\frac{\bigtriangleup^2}{2 M } 
-\frac{c_0 \bigtriangleup^4}{8 M^2} 
-\frac{c_1 \sigma B } { 2 M } ) \psi
\label{NRQCDL}
\end{equation}
where $c_0$ and $c_1$ are coefficients obtained by a
perturbative matching procedure to QCD.
In table~\ref{hybridRESHEAVY} the results of all the recent
NRQCD simulations of the $\overline{b}bg$ hybrids in the quenched
approximation are compiled.  In the
hybrid meson simulations no spin terms are included in the Lagrangian
($c_1 = 0 $ in eq.~\ref{NRQCDL}), so the $1^{-+}$, $0^{+-}$, and $2^{+-}$
states are degenerate. Both the results from the CP-PACS
collaboration~\cite{Manke:1998qc} and from Juge, Kuti and
Morningstar~\cite{Juge:1999ie} were shown to 
be independent of the lattice spacing and lattice volume.
For example, the CP-PACS
collaboration~\cite{Manke:1998qc} found that the masses of the hybrids
were independent of the box size above
$1.2 \; fm$.


The ``asymmetric'' comment in table~\ref{hybridRESHEAVY} refers to the
technique of treating space and time asymmetrically. A smaller lattice
spacing was used in the time direction than in the space direction,
which allowed the signal to be seen for further, for a given spatial
volume.  This technique has helped to reduce the errors. The
practicalities of this idea were demonstrated by Morningstar and
Peardon~\cite{Morningstar:1997ff} for the glueball spectrum.

\begin{table}
\centering
\caption{ \it Mass splitting between the \protect{$\overline{b}b$}
 $1^{-+}$ hybrid and the \protect{$\overline{b}b$} 1S state.
}
\vskip 0.1 in
\begin{tabular}{|c|c|c|} \hline
Group          &  comments & mass GeV \\
\hline
\hline
UKQCD~\cite{Manke:1998yg}   & $O(Mv^4)$ errors  &  $1.68(10)$   \\
CP-PACS~\cite{Manke:1998qc} &  Asymmetric, $O(Mv^4)$ errors  & $1.542(8)$ \\
Juge et al.~\cite{Juge:1999ie}  &  Asymmetric,  $O(Mv^4)$ errors &  $ 1.49(2)(5) $   \\
\hline
\end{tabular}
\label{hybridRESHEAVY}
\end{table}


In table~\ref{hybridRESCHARM} the results of the mass splitting of the
$1^{-+}$ states and the $1S$ state are shown in the charmonium system.
The MILC collaboration used
the standard Wilson and clover actions 
to simulate the charm quark in their  simulations
of heavy exotic mesons, as previous work
has shown this to be reliable~\cite{Davies:1997hv}.

\begin{table}
\centering
\caption{ \it Mass splitting between the \protect{$\overline{c}c$}
 $1^{-+}$ hybrid and the \protect{$\overline{c}c$} 1S state.
}
\vskip 0.1 in
\begin{tabular}{|c|c|c|} \hline
Group          &  comments & mass MeV \\
\hline
\hline
MILC~\cite{Bernard:1997ib,McNeile:1998cp}   & Wilson action  &  $1340^{+60}_{-150}+sys$   \\
MILC~\cite{McNeile:1998cp}   & clover action  &  $1220^{+110}_{-190}+sys$   \\
CP-PACS~\cite{Manke:1998qc} &  NRQCD $O(Mv^4)$ errors, Asymmetric  & $1323(13)$ \\
\hline
\end{tabular}
\label{hybridRESCHARM}
\end{table}



The first results for heavy exotic~\cite{Perantonis:1990dy} hybrids
were done in the adiabatic surfaces approach, where the effect of the
excited glue is subsumed in a potential measured on the lattice
(see~\cite{Kuti:1998rh,Michael:1998sm} for a review).  Juge, Kuti and
Morningstar~\cite{Juge:1999ie} have completed a systematic study of
these potentials. The NRQCD approach is a more accurate approximation to
QCD than the adiabatic potential technique; however Juge, Kuti
and Morningstar~\cite{Juge:1999ie} found that the adiabatic potential
approach reproduced the level splittings from NRQCD up to 10 \%.  A
preliminary result for the calculation of the adiabatic surfaces with
the effects of dynamical fermions included has been reported by 
Bali~\cite{Kuti:1998rh} and collaborators.
No dramatic differences between the quenched theory
result were observed.

\section{Conclusions}

All the lattice simulations agree that the light $1^{-+}$ state has a
mass of $2.0(2) GeV$. The first simulation that included the effects of
dynamical fermions has not changed the result.

The experimental results for light exotics are reviewed by
S.U. Chung~\cite{Chung:1999}, so I just briefly compare the 
lattice results to experiment.
There is an experimental signal for a $1^{-+}$ state at $1.4$ GeV with
a decay into $\eta\pi$~\cite{Chung:1999} from E852, Crystal barrel,
VES, and KEK.  It is surprising that this state is only seen in the
$\eta\pi$ channel, as this decay is theoretically suppressed relative
to other decays~\cite{Page:1997rj}.  The E852 collaboration have also
reported~\cite{Adams:1998ff} a signal for $1^{-+}$ state decaying into
$\rho\pi$ with a mass of around $1.6$ GeV.  The decay width is in
reasonable agreement with theoretical calculations~\cite{Page:1997xs}. 

Clearly the agreement between the possible experimental signals for
the $1^{-+}$ states and the lattice results is very poor.
The errors on the lattice results for $1^{-+}$
states are large relative to the errors on $\overline{Q}Q$ states. To
quantify the disagreement between experiment and lattice results the
systematic errors on the lattice simulation results should be reduced.
In particular the masses of the quarks used in the lattice simulation
should be reduced.  It is possible that the states seen experimentally
are really $\overline{Q}\overline{Q}QQ$ states, in which case the
operators used in the lattice simulations would not couple strongly to
them.

Although the adiabatic lattice potential approach is not expected to
be a good description of the physics of light hybrids, we note that
the results of Juge, Kuti and Morningstar~\cite{Juge:1999ie}, show
that the splitting between the ground and first excited state is about
$200$ MeV, in broad agreement with the experimental results of
the E852 collaboration. Although no insight is gained about the different
decay widths.


To definitely identify an exotic hybrid meson requires both the
calculation of the mass as well as the decay widths. 
There has been
very little work on hadronic decays on the lattice. 
The most obvious hadronic process to study using lattice gauge theory is 
the $\rho \rightarrow \pi\pi$ decay, however there have 
only been a
few attempts to calculate the $g_{\rho \pi \pi}$
coupling~\cite{Gottlieb:1984rh} The GF11 lattice group
has recently computed the decay widths for the decay of the $0^{++}$
glueball to two pseudoscalars~\cite{Sexton:1995kd}.

The MILC collaboration~\cite{Bernard:1997ib} have started
to investigate the 
mixing between the operator in eq.~\ref{eq:interponemp}  and
the operator ($\pi \otimes a_1$) eq.~\ref{eq:QQQQ}.
\begin{equation}
\overline{\psi}^a \gamma_5 \psi^{a} 
\overline{\psi}^b \gamma_5 \gamma_i \psi^{b} 
\label{eq:QQQQ}
\end{equation}
which has the quantum numbers $1^{-+}$.  This type of correlator would
naively be expected to yield the decay width of the $1^{-+}$ state to
$\rho$, and $a_1$.  Unfortunately the analysis of Maiani and
Testa~\cite{Maiani:1990ca} shows that the matrix element required in
the computation of the decay width is hidden beneath an unphysical
term that increases exponentially with time.  The origin of the
unphysical term comes from the requirement that both final mesons
should be onshell and is deeply related to theory being defined in
Euclidean space (required for us to have a well defined theory to
simulate).  Some information may be extracted for onshell
processes~\cite{Maiani:1990ca,Sexton:1995kd}, using the methods
proposed by Michael~\cite{Michael:1989mf}.

\section{Acknowledgements}
I thank Chris Michael and Doug Toussaint for 
discussions about exotic mesons.



\end{document}